\documentclass{appolb}
\usepackage{graphicx}
\usepackage{cite}
\usepackage{hyperref}
\usepackage{amsmath}
\usepackage{graphicx}
\usepackage{dcolumn}
\usepackage{bm}
\usepackage{enumitem, kantlipsum}
\usepackage{graphicx}

\begin{document}
	
\title{A new and finite family of solutions of hydrodynamics: \\ 
 Part III: Advanced estimate of the life-time parameter}
\author{{T. Cs\"org\H{o}$^{1,2}$ and G. Kasza$^{1,2}$,}\\[1ex]
	$^1$EKU KRC, H-3200 Gy\"ongy\"os, M\'atrai \'ut 36, Hungary,\\
	$^2$Wigner RCP, H - 1525 Budapest 114, P.O.Box 49, Hungary\\
}	
	
\maketitle
	
\begin{abstract} 
We derive a new formula for the longitudinal HBT-radius of the two particle Bose-Einstein correlation function from a new family of finite and exact, accelerating  solution of relativistic perfect fluid hydrodynamics for a temperature independent speed of sound. 
The new result generalizes the Makhlin-Sinyukov and Herrmann-Bertsch  formulae and leads to an advanced life-time estimate of  high energy heavy ion and proton-proton collisions.
\end{abstract}

\section{Introduction}
This manuscript is the third part of a manuscript series. This series
 presents various applications of a new, accelerating, finite and
exact family of solutions of perfect fluid hydrodynamics, the recently found Cs\"org\H{o} - Kasza - Csan\'ad - Jiang (CKCJ) family of solution of ref.~\cite{Csorgo:2018pxh}. 
The first part of this series~\cite{Csorgo:2018fbz} fixes the notation, summarizes this class of exact solutions  and evaluates the rapidity and pseudorapidity density
distributions. The second part~\cite{Kasza:2018jtu} evaluates the initial energy densities in high energy collisions~\cite{Csorgo:2018pxh}, and provides a fundamental correction to the renowned Bjorken estimate of initial energy density~\cite{Bjorken:1982qr}.

In this manuscript, we evaluate the Bose-Einstein correlation functions in a Gaussian approximation from the CKCJ solutions~\cite{Csorgo:2018pxh}. Given that the considered dynamics is a 1+1 dimensional expansion, we  evaluate $R_{L}$, the Hanbury Brown - Twiss  (HBT) radii in the longitudinal (beam) direction.  
This longitudinal HBT radius parameter is proportional to the mean freeze-out time of the fireball,
thus the advanced evaluation of its transverse mass dependence and its constant of proportionality for finite, longitudinally non-boost-invariant
fireballs may have important physics implications on life-time determinations.

\vfill\eject

\section{Bose-Einstein correlations and the longitudinal HBT radii}
In high energy heavy ion collisions, Bose-Einstein correlation functions (BECF) measure characteristic sizes of the particle emitting source,
corresponding to lengths of homogeneity~\cite{Makhlin:1987gm}. In high energy heavy ion collisions,  the particle emitting source can be approximated as
a locally thermalized fireball, surrounded by a halo of long-lived resonances, this is the so-called core-halo picture~
The momentum dependent intercept parameter $\lambda_*$ of the two-particle Bose-Einstein correlation function can be
interpreted in the core-halo picture of 
ref.~\cite{Csorgo:1999sj} as follows:
\begin{equation}
\lambda_*=\left(\frac{N_{c}}{N}\right)^2 \, = \left(\frac{N_{c}}{N_c + N_{h}}\right)^2
\end{equation}
where $N = N_{c}+N_{h}$ is the total number of the emitted particles with a given momentum, adding the contributions from both the core
$N_{c}$ and the halo, $N_{h}$. The fireball that undergoes a hydrodynamical evolution corresponds to core~\cite{Csorgo:1999sj}.
For locally thermalized sources, the lengths of homogeneity are expressible in terms of the derivatives of the fugacity, $\exp\left(\mu(x)/T(x)\right)$ and the 
locally thermalized momentum distribution, $\exp\left(- k^\mu u_\mu(x)/T(x)\right)$, 
corresponding to the so called geometrical and thermal length scales~\cite{Csorgo:1999sj}. 
Assuming an effective Gaussian source for the core particles, the BECF can be expressed in terms of the Bertsch-Pratt variables as follows:
\begin{equation}
C(\Delta k, K)=1+\lambda_*\exp\left(-R^2_{side}Q^2_{side}-R^2_{out}Q^2_{out}-R^2_{L}Q^2_{L}-2R^2_{out,L}Q_{out}Q_{L}\right).
\end{equation}
All the fit parameters 
($\lambda_*$, $R_{side}$, $R_{out}$, $R_{L}$ and $R^2_{out,L}$) depend
on the mean momentum of the particle pair, $K^\mu = 0.5 (k_1^\mu+k_2^\mu)$. The four-momentum of a given particle is denoted by $k = (E_k,\textbf{k}) = (E_k, k_x, k_y, k_z)$.
The three-components of the relative and mean momenta are denoted as
\begin{align}
\Delta\textbf{k}&=\textbf{k}_1-\textbf{k}_2,\\
\textbf{K}&=0.5 \left(\textbf{k}_1+\textbf{k}_2\right).
\end{align}
In the Bertsch-Pratt decomposition of the relative momentum~\cite{Bertsch:1989vn,Pratt:1984su}, 
the principal directions are defined as follows: 
The $out$ direction is perpendicular to the beam axis and parallel to the mean 
transverse momentum of the boson pair; 
the longitudinal direction (indicated by subscript $L$) is parallel to the beam axis ($r_z$), and the $side$ direction is orthogonal to the previous two directions.  This Bertsch-Pratt decomposition of the relative momentum is defined as follows:
\begin{align}
	Q_{side}&=\frac{|\Delta \textbf{k}\times \textbf{K}|}{|\textbf{K}|}\\
	Q_{out}&=\frac{\Delta \textbf{k}\cdot\textbf{K}}{|\textbf{K}|}\\
	Q_L&=k_{1,z}-k_{2,z},
\end{align}
If the Bose-Einstein correlation function is an approximately Gaussian in terms of the relative momenta,
the Gaussian HBT radii $R^2_{i,j}$ can be introduced, with $\left\{i,j\right\}  \epsilon \left\{side, out, long\right\}$.
These Gaussian Bertsch-Pratt-radii can be related to the variances of the hydrodynamically evolving core,
while the halo of the long-lived resonances is responsible for the effective reduction of the strength of the correlation function:
\begin{equation}
 R_{i,j}^2=\langle \tilde{x}_i \tilde{x}_j \rangle_c - \langle \tilde{x}_i\rangle_c\langle \tilde{x}_j\rangle_c.
\end{equation}
Here the $\langle A \rangle _c$ stands for the average of quantity $A$ in the core, 
$i,j$ stand for directions (side, out or long) and 
\begin{align}
\tilde{x}_i&=x_i - \beta_i t,\\
\beta_i&=\frac{{k}_{i,1}+k_{i,2}}{E_1+E_2}.
\end{align}
In this manuscript, we focus on the longitudinal radius,
so the radii of the $side$ and $out$ direction are not discussed, see e.g. ref.~\cite{Csorgo:1999sj} for more details on this point.
\begin{figure}[h!]
	\centering
	\includegraphics[scale=0.9]{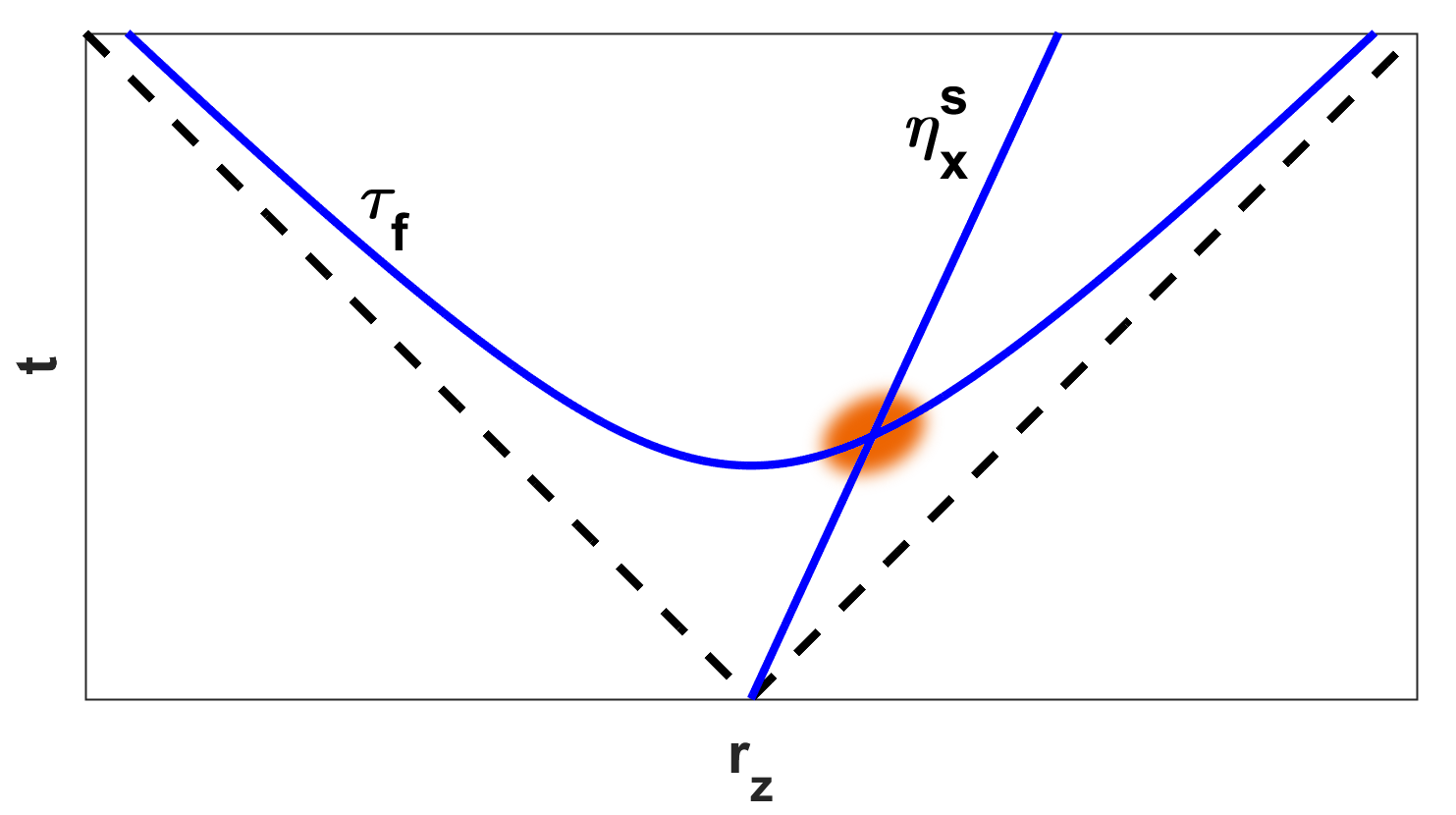}
	\caption{Space-time picture of particle emission for longitudinally expanding fireballs.}
	\label{fig:saddlepoint}
\end{figure}
As discussed in \cite{Csorgo:1995bi}, for a 1+1 dimensional relativistic source, the longitudinal radius in an arbitrary frame reads as
\begin{equation}
R^2_L=\left(\beta_L\sinh(\eta_x^s)-\cosh(\eta_x^s)\right)^2\tau_f^2\Delta \eta_x^2 + \left(\beta_L\cosh(\eta_x^s)-\sinh(\eta_x^s)\right)^2 \Delta \tau^2,
\end{equation}
where $\eta_x$ is the space-time rapidity, and $\eta_x^s$ is the main emission region of the source, which derived by the saddle-point calculation of the rapidity density, $\Delta \tau$ and $\Delta \eta_x$ are characteristic sizes around $\tau_f$ and $\eta_x^s$. This formula simplifies a lot in the LCMS 
(longitudinally co-moving system) frame of the boson pair, where $\beta_L=0$:
\begin{equation}
R^2_L=\cosh^2(\eta_x^s)\tau_f^2\Delta \eta_x^2 + \sinh^2(\eta_x^s) \Delta \tau^2.
\end{equation} 
Our new family of solutions are finite, and limited to a narrow rapidity interval around midrapidity \cite{Csorgo:2018pxh}. 
At mid-rapidity, if $\eta_x^s \approx 0$, the above equation can be simplied even further:
\begin{equation}
R_L=\tau_f \Delta \eta_x.
\end{equation}


\section{Previous results on the longitudinal HBT-radius}

For a Hwa-Bjorken type of accelerationless, longitudinal flow ~\cite{Hwa:1974gn,Bjorken:1982qr}
Makhlin and Sinyukov determined the longitudinal length of homogeneity in ref.~\cite{Makhlin:1987gm} 
as
\begin{equation}\label{MS-formula}
R_{L}=\tau_{Bj}\sqrt{\frac{T_f}{m_T}}.
\end{equation}
In this equation, $T_f$ stands for the freeze-out temperature, $m_T$ is the transverse mass of the particle pair
and $\tau_{Bj}$ is the mean freeze-out time of the Hwa-Bjorken solution. 
This result makes it possible to determine the life-time, i.e. $\tau_{Bj}$ of the reaction from the measurement of the longitudinal HBT radius parameter, provided that $T_f\approx m_{\pi}\approx 140$ MeV can be estimated from the analysis of the single particle spectra.

Evaluating the HBT radii from the same Hwa-Bjorken solution~\cite{Hwa:1974gn,Bjorken:1982qr}, 
Herrmann and Bertsch obtained a more accurate result in ref.~\cite{Herrmann:1994rr}, using a Gaussain approximation  for the longitudinal HBT radius at midrapidity, in terms of Bessel functions $K_1(z)$ and $K_2(z)$, as follows:
\begin{equation}
R_{L}=\tau_f\sqrt{\frac{T_f}{m_T}}\sqrt{\frac{2K_2(m_T/T_f)}{K_1(m_T/T_f)}}.
\end{equation}
This formula improves the Sinyukov-Makhlin formula \eqref{MS-formula} for lower $m_T/T_f$ values, and approaches it in the large $m_T/T_f$ limit.

If the flow is accelerating, the estimated origin of the trajectiories is shifted back in proper-time, thus $\tau_{Bj}$ is underestimating the life-time of the reaction. The correction was estimated, based on the modification of the flow-profile, from the Cs\"org\H o-Nagy-Csan\'ad (CNC) solution \cite{Nagy:2007xn} 
as follows
\begin{equation}
R_{L}=\frac{\tau_f}{\lambda}\sqrt{\frac{T_f}{m_T}},
\end{equation}
where $\tau_f$ stands for the freeze-out time. 
In the $\lambda \rightarrow 1$ boost-invariant limit, this formula also reproduces the Makhlin-Sinyukov formula,
but for the realistic $\lambda > 1 $ parameter values it yields larger life-times as compared to the Makhlin-Sinyukov formula.

\section{The longitudinal HBT-radius parameter of the CKCJ solution}
Let us evaluate  the emission function for the CKCJ solution of refs.~\cite{Csorgo:2018pxh,Csorgo:2018fbz,Kasza:2018jtu}. 
The integration of the Cooper-Frye formula is performed by the saddle-point approximation. 
Near to mid-rapidity, the fluid rapidity is well approximated by a linear function of the space-time rapidity: 
$\Omega \approx \lambda \eta_x$. Using  a saddle-point integration in $\eta_x$, we  obtain 
the  rapidity distribution:
\begin{equation}
\frac{dN}{dy}\approx\frac{\left(2 \pi \Delta \eta_x^2\right)^{1/2}}{2\pi\hbar} \left[ k_{\mu}u^{\mu} \frac{\tau(\eta_x)}{\cosh(\Omega-\eta_x)}\exp\left(-\frac{k_{\mu}u^{\mu}}{T_f(\eta_x)}\right)\right]_{\eta_x=\eta_x^s}.
\end{equation}
Here $\eta_x^s$ stands for the saddle-point, which is found to be proportional to  the rapidity $y$:
$\eta_x^s\approx\frac{y}{2\lambda-1}$.
At midrapidity, the saddle-point vanishes and the emission function can be well  approximated by a Gaussian centered on zero. The width of this Gaussian is given by $\Delta \eta_x$ as 
\begin{equation}
\Delta \eta_x \approx \sqrt{\frac{T_f}{m_t}}\frac{1}{\sqrt{\lambda(2\lambda-1)}}.
\end{equation}
At mid-rapidity, these considerations lead to the following longitudinal HBT-radius parameter:
\begin{equation}
R_L=\tau_f \Delta \eta_x \approx \frac{\tau_f}{\sqrt{\lambda\left(2\lambda-1\right)}} \sqrt{\frac{T_f}{m_T}}.
\label{e:Rlong-ckcj}
\end{equation}
Surprisingly,  this result is independent of the equation of state, and it is formally different from the CNC estimate.

\begin{figure}[h!]
	\centering
	\includegraphics[scale=0.42]{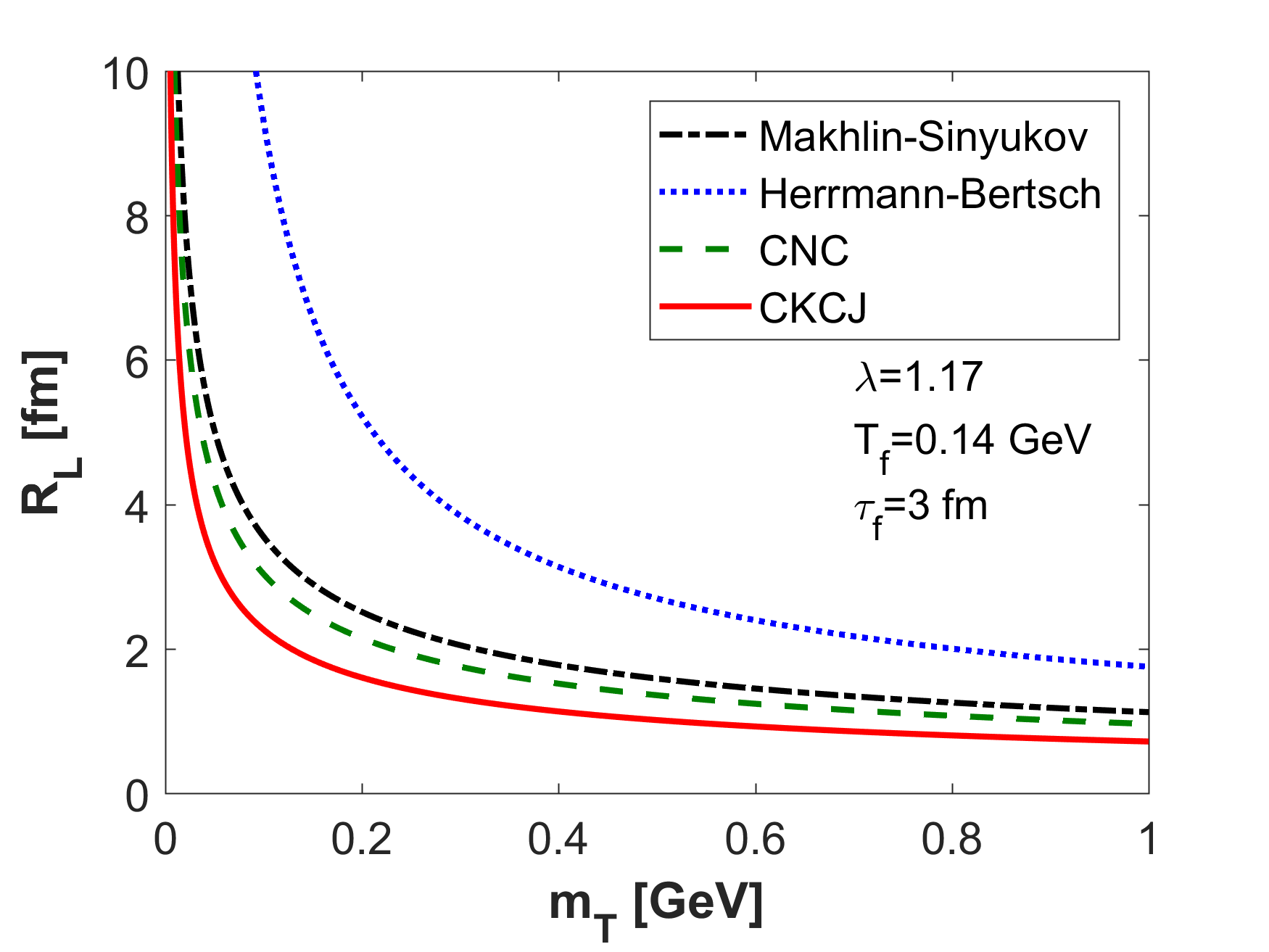}
	\includegraphics[scale=0.42]{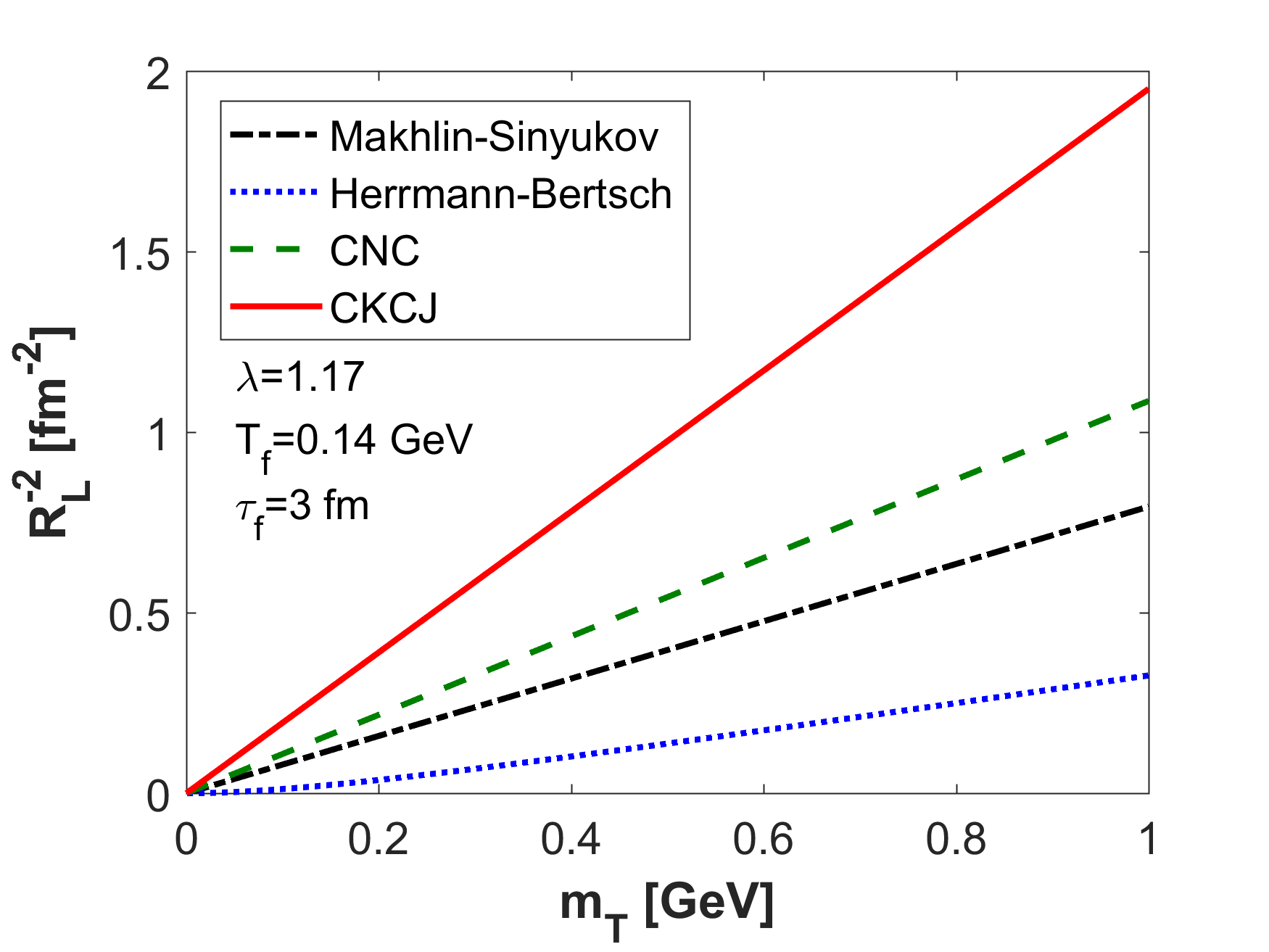}
	\caption{The HBT radius $R_L(m_T)$ (left) and $1/R_{L}^2(m_T)$ (right) of the CKCJ solution are shown with solid red lines and compared to earlier estimations. The parameters correspond to fit results of the CKCJ solution to p+p collisions at $\sqrt{s}=7$ TeV~\cite{Csorgo:2018fbz}.}
	\label{fig:rlong}
\end{figure}

Our result thus presents and important step forward: once the parameter $\lambda$ of the acceleration is determined from the
fits to the (pseudo)rapidity distributions~\cite{Csorgo:2018fbz}, this parameter combined with the longitudinal HBT radius measurement can be used to
provide an advanced estimate of the life-time of the reaction, solving eq. ~(\ref{e:Rlong-ckcj}) for the life-time $\tau_f$.
The significance of our advanced formula is illustrated
on Figure~\ref{fig:rlong}.

\section*{Acknowledgments}
We greatfully acknowledge partial
support form the Hungarian NKIFH grants No. FK-123842 and FK-123959, the Hungarian
EFOP 3.6.1-16-2016-00001 project and the exchange programme of the Hungarian and the Ukrainian
Academies of Sciences, grants NKM-82/2016 and NKM-92/2017.

\end{document}